\documentclass[12pt]{iopart}
\usepackage{graphicx}

\newcommand\lapp{\mathrel{\rlap{\lower4pt\hbox{\hskip1pt$\sim$}}
    \raise1pt\hbox{$<$}}}
\newcommand\gapp{\mathrel{\rlap{\lower4pt\hbox{\hskip1pt$\sim$}}
    \raise1pt\hbox{$>$}}}
\newcommand\eapp{\mathrel{\rlap{\raise2pt\hbox{\hskip0pt$\sim$}}
    \lower1pt\hbox{$-$}}}

\begin{document}
%%%%%%%%%%%%%%%%%%%%%%%%%%%%%%%%%%%%%%%%%%%%%%%%%%%
\newcommand{\be}{\begin{equation}}
\newcommand{\ee}{\end{equation}}
\newcommand{\ben}{\begin{eqnarray}}
\newcommand{\een}{\end{eqnarray}}
\newcommand{\bb}{\bibitem}
\newcommand{\ov}{\overline}
\newcommand{\wt}{\widetilde}
\newcommand{\nn}{\nonumber}
%%%%%%%%%%%%%%%%%%%%%%%%%%%%%%%%%%%%%%%%%%%%%%%%%%%

\title[Cosmic strings in Bekenstein-type models]{Cosmic strings in Bekenstein-type models}

\author{J Menezes\dag, PP Avelino \dag, and C Santos \dag}

\address{\dag\ Centro de F\'{\i}sica do Porto e Departamento de F\'{\i}sica
da Faculdade de Ci\^encias da Universidade do Porto, Rua do Campo Alegre 687,
4169-007, Porto, Portugal}

\begin{abstract}
We study static cosmic string solutions in the context of 
Bekenstein-type models. 
We show that there is a class of models of this type for which 
the classical Nielsen-Olesen vortex is still a valid solution. However, 
in general static string solutions in Bekenstein-type models strongly 
depart from the standard Nielsen-Olesen solution with the electromagnetic 
energy concentrated along the string core seeding spatial variations of 
the fine structure constant, $\alpha$.
We consider models with a generic gauge kinetic function and show that
equivalence principle constraints impose tight limits on the allowed
variations of $\alpha$ induced by string networks on cosmological scales.
\end{abstract}
\ead{jmenezes@fc.up.pt}
\maketitle

\section{Introduction}

Cosmological theories motivated by models with extra-spatial 
dimensions \cite{Polchinski} have recently attracted much attention, 
in large measure due to the 
possibility of space-time variations of the so-called $``$constants$"$ 
of Nature \cite{barrow,Uzan,VFC,Essay}. Among these are 
varying-$\alpha$ models  
(where $\alpha$ is the fine-structure constant) such as the one proposed 
by Bekenstein \cite{Bekenstein}. In Bekenstein-type models  
the variation of the fine structure constant, 
$\alpha=e^2/4\pi\,{\hbar}\,c$, is sourced by a scalar field, $\varphi$,  
coupling minimally to the metric and to the electromagnetic term, 
$``$$F^2$$"$, by a non-trivial gauge kinetic function, $B_F(\varphi)$.
The interest in this type of models has been increased with recent results 
coming from both quasar absorption systems \cite{Webb,Murphy} 
(see however \cite{Chand,srianand}) and the Oklo natural nuclear 
reactor \cite{Fujii} 
suggesting a cosmological 
variation of $\alpha$ at low red-shifts. Other constraints at low redshift 
include atomic clocks \cite{Marion} and meteorites \cite{Olive}. At high 
redshifts there are also upper limits to the allowed variations of $\alpha$ 
coming from either the Cosmic Microwave Background or 
Big Bang Nucleossynthesis 
\cite{hannestad,Avelino,Martins,Rocha,wmap1,sigurdson,wmap2}.
  
Meanwhile on the theoretical side some effort has been made on the 
construction of  
models which can explain a variation of $\alpha$ at redshifts $z \sim 1-3$ 
of the same order as that reported in refs. \cite{Webb,Murphy} 
while being consistent with the other constraints at lower redshifts. 
Although this cannot be achieved in the simplest class of models in 
which the potential and gauge kinetic function are expanded up to linear 
order \cite{linear} there are other classes of models which 
seem in better agreement with the data (see 
for example \cite{Lee}). 
Moreover, in some of these models 
\cite{linear,Lee,OlivePos,Anchordoqui,Parkinson,Copeland,Nunes} 
the fine structure constant is directly 
related to the scalar field responsible for the recent acceleration of 
the Universe \cite{Tonry,Bennett}.

A number of authors have also studied the spatial variations of the fine 
structure constant induced by fluctuations in the matter fields showing 
that they are proportional to the gravitational potential and are typically 
very small to be detected directly with present day technology except 
perhaps in the vicinity of compact objects with strong gravitational 
fields (see for example \cite{Sandvik,mota1,mota2,mota3}).

The variation of $\alpha$ in the early Universe has even been associated 
with the solutions to some of the problems of the standard cosmological 
model \cite{vslcos,vslcos1}. 
However, it has been shown that the ability of specific models to solve 
some of the problems of the standard cosmology (in particular the horizon 
and flatness problems) is directly related to the evolution of 
`cosmic numbers' which are dimensionless parameters involving cosmological 
quantities rather than the evolution of dimensionless combinations 
of the so-called `fundamental constants of nature' 
\cite{vsl,cosmic}.

Note that a change in the value of the fine structure constant should in 
principle be accompanied by a variation of other fundamental constants as 
well as the grand unification scale. However, we shall adopt 
a phenomenological approach and neglect possible variations of other 
fundamental constants. 

In this article we consider the case of 
static Abelian vortex solutions whose electromagnetic energy 
is localized along a stable string-like core which acts as a source for 
spatial variations of $\alpha$ in the vicinity of the string. We generalize 
to a generic gauge kinetic function the work of ref. \cite{Magueijo} and study 
the limits imposed by the Weak Equivalence Principle \cite{Will,Damour} 
on the allowed 
cosmological variations of $\alpha$. The article is organized as 
follows. In Sec. II we briefly introduce Bekenstein-type models and obtain the 
equations describing a static string solution. We describe the numerical 
results obtained for a number of possible choices of the gauge kinetic 
function in Sec. III discussing the possible cosmological implications of 
string networks of this type in the light of equivalence principle 
constraints in Sec. IV. Finally, in Sec. V 
we summarize 
our results and briefly discuss further prospects. 
Throughout this paper we shall use units in which $\hbar = c = 1$ 
and the metric signature $+---$.

\section{Bekenstein-type models}

We first review Bekenstein-type models with a charged
complex scalar field $\phi$ undergoing spontaneous symmetry breaking.
Let $\phi$ be a complex scalar field with a U(1) gauge symmetry and $a_\mu$
be the gauge field. Let us also assume that the electric charge is a function
of space and time coordinates, $e=e_0 \epsilon(x^\mu)$ where $\epsilon$ is
a real scalar field and $e_0$ is an arbitrary constant charge.
The Lagrangian density in Bekenstein-type models is given by 
(see for example \cite{Magueijo}):
\ben
{\cal L}\, &=& \left(D_\mu\,\phi\right)^{*}\left(D^\mu\,\phi\right)-V(\phi)\,
-\frac{1}{4}\,B_F(\varphi)\,f_{\mu\nu}\,f^{\mu\nu} \nn \\
&+&\frac{1}{2}\,\partial_\mu\,\varphi\,\partial^\mu
\,\varphi \, , \label{lll}
\een
where $B_F(\varphi)=\epsilon^{-2}(\varphi)$ is the gauge kinetic function and 
$\varphi$ is a scalar field.  
In eqn. (\ref{lll}),
$D_\mu\,\phi=\left(\partial_\mu-i e_0 a_\mu\right)\,\phi$ are covariant
derivatives and the electromagnetic field tensor is given by
\be
f_{\mu\nu}\,=\,\partial_\mu  a_\nu -\partial_\nu a_\mu \, .
\ee
The function $B_F=\epsilon^{-2}(\varphi)$ acts as the effective dielectric 
permittivity 
which can be phenomenologically taken to be an arbitrary function of $\varphi$.
We assume that $V(\phi)$ is the usual Mexican hat potential with
\be
V(\phi)\,=\,\lambda\left(|\phi|^2+\frac{m^2}{2\lambda}\right)^2 \, ,
\ee
where $\lambda>0$ and $m^2 < 0$ are constant parameters and 
$\eta = m/\sqrt{2} \lambda$ is the symmetry breaking scale.
The Lagrangian density in eqn. (\ref{lll}) is then invariant 
under U(1) gauge transformations of the form 
$\delta\, \phi=-i\zeta\,\phi$, $a_\mu \, \rightarrow\, a_\mu+
\zeta_{,\mu}$.

Zero variation of the action with respect to the complex conjugate of
$\phi$, i.e., $\phi^\star$, gives:
\be\label{eqphi}
D_\mu\,D^\mu\,\phi\,=\,-\frac{\partial\, V}{\partial\,\phi^\star}\,.
\ee
Variation with respect to $a_\mu$ leads to:
\be\label{eqfmn}
\partial_\nu\left[B_F(\varphi)\,f^{\mu\nu}\right]\,=\,j^\mu\,,
\ee
with the current $j^\mu$ defined as
\be
j^\mu = i\,e_0\,\left[\phi\,\left(D^\mu\,\phi\right)^\star
-\phi^\star\,\left(D^\mu\,\phi\right)\right].
\ee
Finally,  variation with respect to $\varphi$ gives
\be\label{eqbox}
\partial_\mu\,\partial^\mu\,\varphi\,=\,-\frac{1}{4}\,
\frac{\partial B_F(\varphi)}{\partial\, \varphi}\, f^2\,.
\ee

We now look for the static vortex solutions in these theories adopting the
following ansatz
\ben
\phi &=& \chi(r)\,e^{i n \theta}\, , \label{n}\\
a_\theta  &=& a(r)\, , \label{bansatz}
\een
with all other components of $a_\mu$ set to zero. Here we are using
cylindrical coordinates ($r,\theta,z$), $n$ is the winding
number and $\chi(r)$ is a real function of $r$.
Substituting the ansatz given in (\ref{n}-\ref{bansatz}) into 
eqns. (\ref{eqphi}-\ref{eqbox}) one gets 
\ben
&&\frac{1}{r}\frac{d}{dr}\left(r\frac{d\chi}{dr}\right)-
\left[\left(\frac{n}{r}-e_0\,a\right)^2
+m^2+2\lambda \chi^2\right]\chi=0, \label{pri}\\
&&\frac{d}{dr}\left(B_F\frac{1}{r}\frac{d}{dr} \left(r a\right)\right)
+2e_0\left(\frac{n}{r}-e_0a\right)\chi^2=0,\label{seg}\\
\label{eqvarphi}
&&-\frac{1}{r}\frac{d}{dr}\left(r\frac{d\varphi}{dr}\right)+\frac{1}{2}
\frac{dB_F(\varphi)}{d\varphi}
\left(\frac{1}{r}\frac{d}{dr} \left(ra\right)\right)^2=0.\label{ter}
\een
Note that
\be
f^2 = 2 f^{r\theta} \, f_{r\theta} = 2 \left[ \frac1{r} \frac{d}{dr}(ra) \right]^2
\label{error}
\ee
which substituted in eqn. (\ref{eqbox}) gives the factor of one-half in 
the second term on the left hand side of eqn. (\ref{ter}). This corrects 
an error in ref. \cite{Magueijo} which neglected the factor of $2$ in 
eqn. (\ref{error}).

We also investigate the dependence of the energy density on the radial 
coordinate $r$.
For static strings the stress-energy tensor take a diagonal
form with the energy density of the vortex being
$\rho = T^0_0 = g^{00} T_{00}=T_{00} $,
with
\ben
\rho &=&  \left( \frac{d \chi}{dr}  \right)^2 + \left( \frac{d \varphi}{dr}  \right)^2 
  + \frac{1}{2\,r^2} B_F \left( \frac{dv}{dr}  \right)^2 \nn \\&+&  \left( \frac{n - e_0 v}{r} \right)^2 \chi^2  + \lambda 
\left( \chi^2 + \frac{m^2}{2\lambda}\right)^2,
\een
while the spatial components of the stress-energy tensor are given
by $T^{i}_{j} = diag(p_r, p_\theta, p_z)$ with $p_z = - \rho$ (here $v=a\,r$).
Therefore, the energy density of the
vortex, $\rho$, is everywhere positive, while the longitudinal pressure $p_z$
is negative. In fact, this is also one of the defining
features of canonical cosmic strings.

\section{Numerical Solutions}

We consider solutions to the coupled non-linear equations 
for a static straight string. Since no
exact analytic solution has yet been found it proves useful to reduce
equations (\ref{pri}-\ref{ter}) to a set of first order
differential equations for numerical implementation.

Let us introduce three new variables
\ben
\frac{d\,\chi}{dr}&=& \sigma \, , \label{Ia}\\
\frac{d\,v}{dr}&=& b\,r \, ,\label{IIa}\\
\frac{d\varphi}{dr}&=&\eta \, ,\label{IIIa}
\een
with $v=a\,r$. Then by substituting (\ref{Ia}-\ref{IIIa}) into
equations (\ref{pri}-\ref{ter}), one gets
\ben
\label{IV}
\frac{d\sigma}{dr} &=&-\frac{\sigma}{r}+\left[\left(\frac{n-e_0\,v}{r}\right)^2
+m^2+2\,\lambda\,\chi^2\right]\chi\,, \\
\label{V}
\frac{d\,b}{dr}\, &=&\,\frac{1}{B_F}\,\left[-\frac{dB_F}{d\varphi}\,\eta\,b
-2e_0\left(\frac{n-e_0\,v}{r}\right)\,\chi^2\right]\,, \\
\label{VI}
\frac{d\eta}{dr}\,&=&\,-\frac{\eta}{r}+\frac{1}{2}\,\frac{dB_F}{d\varphi}\,b^2\,.
\een
Then we have, at all, a set of six ordinary first order differential
equations, which requires, at least, six boundary conditions to be solved numerically.
The appropriate boundary conditions are \cite{Magueijo}
\be
\lim_{r \rightarrow 0} \chi(r) = 0\,\,\,\,\,\,\,\,\,\,
\lim_{r \rightarrow \infty} \chi (r) =  \sqrt{\frac{-m^2}{2\lambda}}\, ,
\ee
\be
\lim_{r \rightarrow 0} v (r) = 0\,\,\,\,\,\,\,\,\,\,\,
\lim_{r \rightarrow \infty} v(r) =  \frac{n}{e_0} , 
\ee
\be
\lim_{r \rightarrow 0} B_F (r) = 1\,\,\,\,\,\,\,\,\,\,\,
\lim_{r \rightarrow 0} \eta =  0\, . 
\ee
Therefore we have a two point boundary value problem with four conditions
at the origin and two conditions far from the core. 
Let us start by discussing the boundary conditions far from the core 
($r \to + \infty$). In this limit the scalar field has a constant value 
given by $\chi={\sqrt {-m^2/(2\lambda)}}$ and from equation 
(\ref{pri}) one 
immediately sees that this implies that $v(r)=n/e_0$ far 
from the core. There are also four boundary conditions at the string 
core. In this limit, the phase $\theta$ in equation (\ref{n}) is 
undefined which implies 
that $\chi$ must vanish at $r=0$. Also, $v$ must vanish at the string 
core. Otherwise the magnetic energy density would diverge at the core 
of the string. We also normalize the electric charge such that $e=e_0$ 
at the string core (this gives the boundary condition for $B_F$). Finally, 
the boundary condition for $\eta$ becomes evident by using the Gauss law 
to solve equation 
(\ref{ter}) assuming that there are no sources of $\alpha$ variation 
other than the string.

In order to solve this problem numerically we used the relaxation
method which replaces the set of six ordinary differential equations by
finite-difference equations on a mesh of points covering the range
of the integration. This method is very efficient if a good initial
guess is supplied. In our case, the solutions of the standard Nielsen-Olesen
vortex can be used to generate a good initial guess.
We checked that our code reproduces the results for the standard 
Nielsen-Olesen vortex if $B_F = 1$. Throughout this paper we shall 
take $\lambda = 1/2$ and $n = 1$ for definiteness and use units in 
which $\hbar=c=-m^2=1$.

\subsection{Exponential coupling}

The general prescription detailed above can be particularized to specific
choices of gauge kinetic functions.
First, let us consider
\be
B_F(\varphi)\,=\,e^{-\frac{2\varphi}{\sqrt{\omega}}}\, . \label{expo}
\ee

Then the Lagrangian density (\ref{lll}) becomes
\ben
{\cal L}\,&=&\left(D^\mu\,\phi\right)^{*}\,\left(D_\mu\,\phi\right)-V(\phi)\,-\frac{1}{4}\,
\frac{f^{\mu\nu}\,f_{\mu\nu}}{\epsilon^2}\,+\nn \\
&+& \frac{\omega}{2\,\epsilon^2}\,
\partial_\mu\,\epsilon\,\partial^\mu\,\epsilon \, , \label{llagrangian}
\een
which is now written in terms of the field
\be
\epsilon\,=\,\frac{e}{e_0}\,=\,\exp\left(\frac{\varphi}{\sqrt{\omega}}\right)
\ee
and recovers the original Bekenstein model (studied in detail by Magueijo 
et al in ref. \cite{Magueijo} in a similar context). In this model $\omega$ 
is a coupling constant.

\begin{figure}
\begin{center}
\includegraphics*[width=9cm]{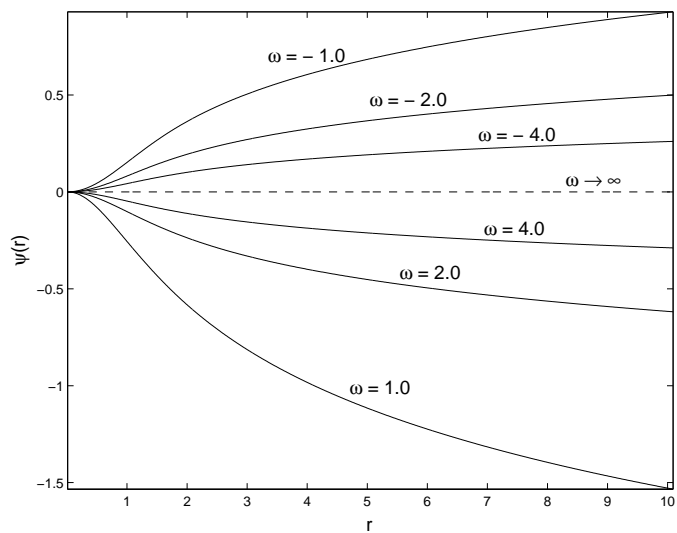}
\end{center}
\caption{The numerical solution of the scalar field 
$\psi \equiv \ln{\epsilon}$ as a function of distance, $r$, 
to the core of string, for the original Bekenstein model. 
If $\omega < 0$ then $\epsilon \to \infty$ when $r \to \infty$.
On the other hand, if $\omega > 0$ then $\epsilon \to 0$ when $r \to \infty$.
The dashed line represents the constant-$\alpha$ theory, which corresponds 
to the limit $ \omega \rightarrow \infty$.} \label{variavel}
\end{figure}

It is easy to show that in the limit $\omega\,\rightarrow\,\infty$ one 
recovers the Nielsen-Olesen vortex with constant $\epsilon$. Although 
the gauge kinetic function in eqn. (\ref{expo}) is only well defined for 
$\omega > 0$ the model described by the Lagrangian density in 
eqn. (\ref{llagrangian}) allows for both negative and positive values 
of $\omega$. 
However, note that if $\omega <0$ the energy density is no longer positive 
definite. 
%We have verified
%the results of ref. \cite{Magueijo} for both positive and negative values of
%$\omega$. 
In  Fig.\ref{variavel} we plot the numerical solution of the scalar field 
$\psi \equiv \ln{\epsilon}$ as a function of distance, $r$, 
to the core of string, in the context of the original Bekenstein model. 
Note that if $\omega < 0$ then $\epsilon$ 
diverges asymptotically away from the string core. On the other 
hand if $\omega > 0$ then $\psi$ goes to zero when $r  \to \infty$. 
In the large $\omega$ limit the curves for positive and negative 
$\omega$ are nearly symmetric approaching the dashed line representing 
the constant-$\alpha$ model when $\omega \rightarrow \infty$.

\subsection{Polynomial coupling}

Another example of a class of gauge kinetic functions is given by
\be
B_F(\varphi)=1.0+\sum_{i=1}^{N}\,\beta_i\,\varphi^i \, ,
\ee
in which $\beta_i$ are dimensionless coupling constants and $N$ is an integer.
If $\beta_1=0$ it is easy to verify that the classical Nielsen-Olesen
vortex solution with constant $\alpha$ is still a valid solution.
This means that there is a class of gauge kinetic functions for which the
classical static solution is maintained despite the modifications to the model.
Substituting  the gauge kinetic function in eqn. (\ref{ter})
one gets
\ben\label{oddeven}
\frac{1}{r}\frac{d}{dr}(r\frac{d\varphi}{dr})&=&
\frac{b^2}{4}\left(\sum_{k=1}^{N}(2k-1)\beta_{2k-1}\varphi^{2k-2}\right)\,
+\nn  \\
&+& \frac{b^2}{4}\left(\sum_{k=1}^{N}(2k)\beta_{2k}\varphi^{2k-1}\right)\,.
\een
which shows that 
the transformation $\beta_i \rightarrow -\beta_i$ for odd $i$
modifies the sign of the solution of $\varphi(r)$ without changing 
$\chi$ or $b$ since $B_F$ is kept invariant.

We will see that both for $\beta_1 \, >\,0$ and $\beta_1 \, <\,0$ the 
behaviour 
of $\psi \equiv \ln \epsilon$  is similar to that of the original Bekenstein 
model described by the lagrangian density in eqn. (\ref{llagrangian}), 
with $\omega\,>\,0$, in particular in the 
limit of small $|\beta_1|$/large $\omega$.
In fact a polynomial expansion of the exponential gauge kinetic function of 
the original Bekenstein model (see eqn. (\ref{expo})) has 
$\beta_1=-2/{\sqrt \omega}$. 
This relation between the
models arises from the fact that the exponential coupling by Bekenstein 
theory can be expanded in a series of powers of $\varphi$ according to 
\be
\beta_i = \frac{(-2)^i}{w^{i/2} i!}\,.
\ee
On the other hand, as mentioned before, if $\beta_1\,=\,0$ one recovers 
the standard result for Nielsen-Olesen vortex with constant-$\alpha$, 
independently of the chosen values of $\beta_i$ for $i > 1$.

We have studied the behaviour of the solutions of eqns. (\ref{Ia}-\ref{VI}) 
for various values of $N$ but for simplicity we shall only consider 
$N \le 2$ in this paper. In particular, we consider 
\be
B_F = 1.0 + \beta_1 \varphi + \beta_2 \varphi^2, \label{nosso}
\ee
with two free parameters.

%Another example of a class of gauge kinetic functions is given by the 
%polynomial expansion,
%\be
%B_F(\varphi) = 1.0 + \sum_{i
%= 1}^{N} \beta_i \varphi^i \label{polin}
%\ee
%in which $\beta_i$
%are dimensionless coupling constants. If $\beta_1=0$ it is easy to verify 
%that the classical Nielsen-Olesen vortex solution with constant $\alpha$ 
%is still a valid solution in this context. This means that there is a 
%class of gauge kinetic functions for which the classical static solution is 
%maintained despite the modifications to the model. It is also easy to show 
%that the transformation $\beta_i \to - \beta_i$ for even $i$ modifies the 
%sign of the solution of $\varphi(r)$ but does not change $\chi$ or $b$.

%We have studied the behaviour of the solutions of eqns. (\ref{I}-\ref{VI}) 
%for various values of $N$ but for simplicity we shall only consider 
%$N \le 2$ in this paper. In particular, we consider 
%where
%\be
%B_F = 1.0 + \beta_1 \varphi + \beta_2 \varphi^2, \label{nosso}
%\ee
%with two free parameters.

\begin{figure}
\begin{center}
\includegraphics*[width=9cm]{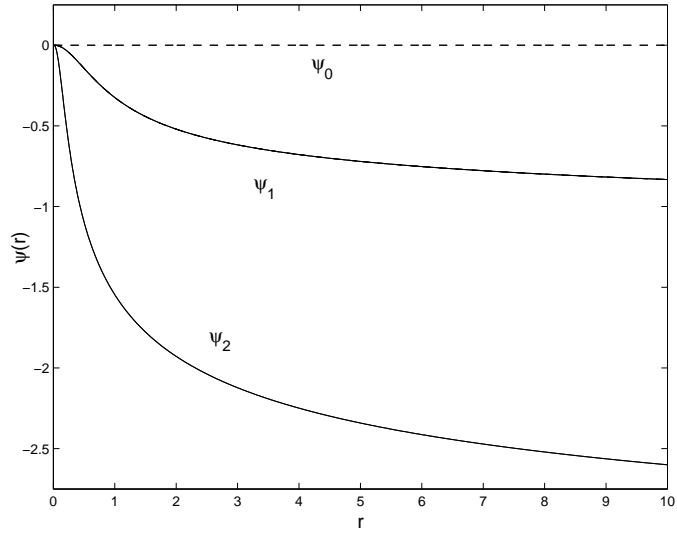}
\end{center}
\caption{The numerical solution of the scalar field 
$\psi \equiv \ln \epsilon$ as a function of distance, $r$, 
to the core of string, for a polynomial gauge kinetic function. 
Models 0, 1 and 2 are defined by $\beta_1=0$ ($\beta_2$ arbitrary),  
$\beta_1=-3,\,\beta_2=0$ (linear coupling) and 
$\beta_1=-5,\,\beta_2=10$ respectively.  
}
\label{bfs}
\end{figure}

%If one considers $\beta_1 \, >\,0$ ($\beta_1 \, <\,0$) we have a similar 
%behaviour to that of the original Bekenstein model 
%described by eqn. (\ref{expo}), 
%with $\omega\,<\,0$ ($\omega\,>\,0$), in particular in the 
%limit of small $\beta_1$/large $\omega$. This relation between the models 
%arises from the fact that the exponential coupling by Bekenstein theory can 
%be expanded in a series of powers of $\varphi$.
%On the other hand, as mentioned before, if $\beta_1\,=\,0$ one recovers 
%the standard result for Nielsen-Olesen vortex with constant-$\alpha$, 
%independent of the chosen values of $\beta_i$ for $i > 1$.

\begin{figure}
\begin{center}
\includegraphics*[width=9cm]{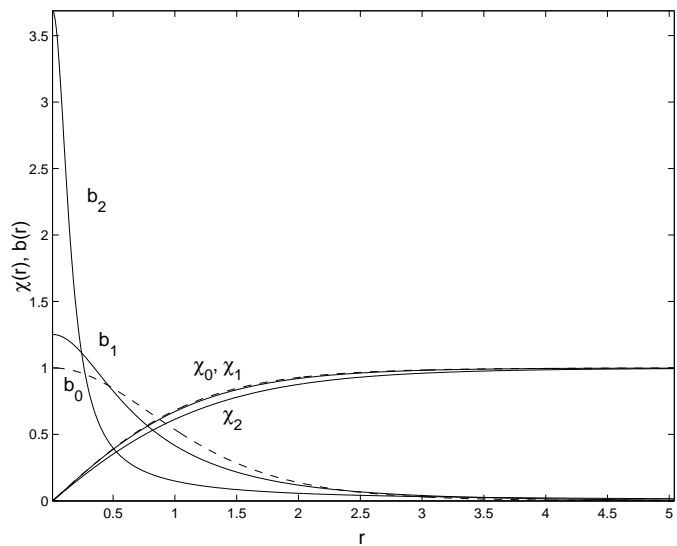}
\end{center}
\caption{
The numerical solution of the fields $\chi(r)$ and $b(r)$ as a function of 
distance, $r$, to the core of string, for models $0$, $1$ and $2$. Note 
that the change in $b(r)$ with respect to the standard 
constant-$\alpha$ result is much more dramatic than the change in $\chi(r)$.}
\label{final1}
\end{figure}

In Fig. \ref{bfs} we plot the numerical solution of the scalar field 
$\psi \equiv \ln \epsilon$ as a function of distance, $r$, 
to the core of string, for a polynomial gauge kinetic function. 
Models $1$ and $2$ are defined $\beta_1=-3,\,\beta_2=0$ (linear coupling) and 
$\beta_1=-5,\,\beta_2=10$ respectively. 
Model 0 (dashed line) represents any model with 
$\beta_1 = 0$ and has $\alpha= {\rm constant}$. Note that the replacement 
$\beta_1 \to -\beta_1$ does not modify the solution for $\psi$.

\begin{figure}
\begin{center}
\includegraphics*[width=9cm]{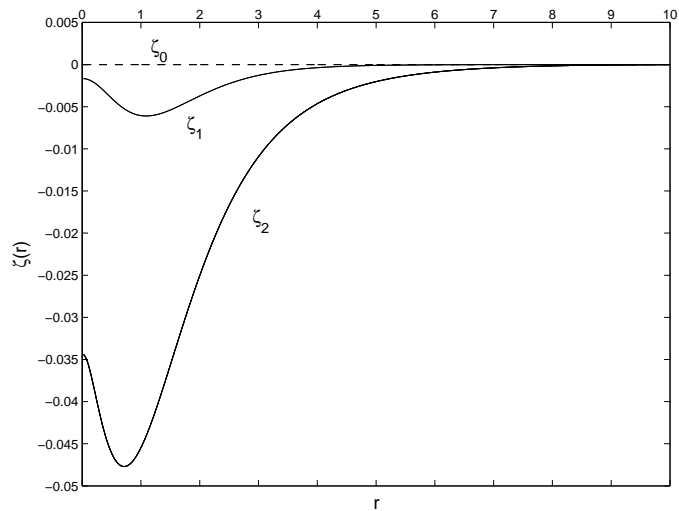}
\end{center}
\caption{Plot of $\zeta_i(r)=\log(\chi_i/\chi_0)$ for the different 
polynomial gauge kinetic functions. Note that although this was not very 
visible in Fig. \ref{final1}, even a small value of 
$\beta_1$ leads to a change of the vortex solution.}
\label{final4}
\end{figure}

In Fig. \ref{final1} we plot the numerical solution of the fields 
$\chi(r)$ and $b(r)$ as a function of distance, $r$, to the core of string, 
for models $0$, $1$ and $2$. We see that the change in $b(r)$ with respect 
to the standard constant-$\alpha$ result is much more dramatic than the 
change in $\chi(r)$. In order to verify the modification to $\chi(r)$ in more 
detail we define the function
\be
\zeta_i(r) = \log\left( \frac{\chi_i}{\chi_0}\right), \label{zeta}
\ee
and plot in Fig. \ref{final4} the results for the different models. 
We clearly 
see that even a small value of $\beta_1$ leads to a modification of the vortex 
solution with respect to the standard Nielsen-Olesen solution.

\begin{figure}
\begin{center}
\includegraphics*[width=9cm]{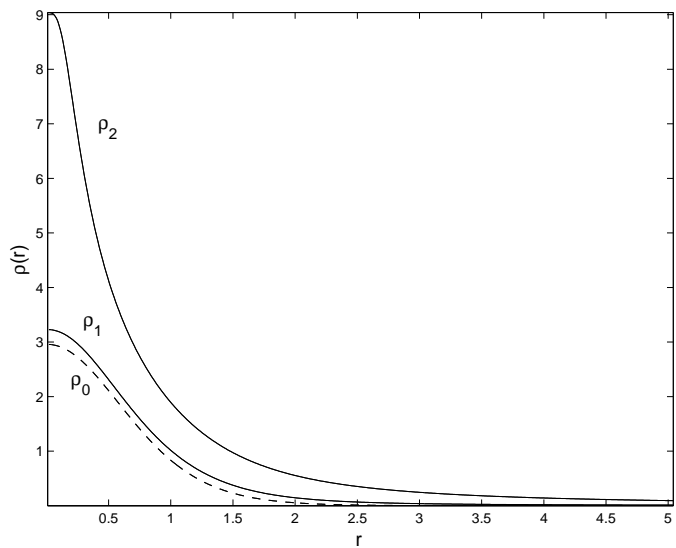}
\end{center}
\caption{The energy density as a function of the 
distance, $r$ to the string core for models 0, 1 and 2. 
The dashed line represents the constant-$\alpha$ model
These results clearly show an increase of the energy density due to the 
contribution of the extra field $\varphi$.}
\label{energy}
\end{figure}

Finally, we have also studied the behaviour of the energy density in this
model. In fact, since the fine structure constant varies we have a new 
contribution due to the field $\varphi$, to the total energy of the 
topological 
defect. As has been previously discussed in ref. \cite{Magueijo} the 
contributions to the energy density of the string can be divided into two 
components. One component is localized around the string 
core (the local string component) the other is related to the 
contribution of the kinetic term associated 
with the spatial variations of the fine structure constant and is not 
localized in the core of the string. The energy profile of this last 
contribution is analogous to that of a global string, 
whose energy per unit length diverges asymptotically far from the core. 
In Fig. \ref{energy} we plot the 
string energy density 
as a function of the distance, $r$ to the string core for models $0$, $1$ 
and $2$. 
The dashed line represents the constant-$\alpha$ model. We clearly see an 
increase of the energy density due to the contribution of the extra field 
$\varphi$.

As an aside, let us point out that the Lagrangian density in (\ref{lll}) 
can be generalized to include a potential $V(\varphi)$. If $V(\varphi)$ has 
a minimum and it is steep enough near it then the spatial variations 
on the value of the fine structure constant can be significantly reduced. 
Also, if 
$V(\varphi)$ has more than one minimum then the Lagrangian density in 
(\ref{lll}) admits domain wall solutions which separate regions with 
different values of the fine structure constant. 
In the absence of an electromagnetic field these solutions reduce to the 
standard ones. On the other hand, if there are sources of electromagnetic 
field present (such as cosmic strings) then there may also be small 
fluctuations on the the value of the fine structure constant generated due 
to the electromagnetic source term in eqn. (\ref{eqbox}). These fluctuations 
are expected to be more localized due to the influence of the potential 
$V(\varphi)$.

\section{Cosmological implications of varying-$\alpha$ strings}

Having discussed in previous sections static string solutions in
the context of Bekenstein-type models it is now interesting to
investigate if such cosmic string networks can induce measurable
space-time variations of $\alpha$ in a cosmological setting. In order
to answer this question we recall that $\varphi$
satisfies the Poisson equation:
\be
\nabla^2 \varphi = \frac{1}{4}\, 
\beta \, f^2\,.\label{poisson}
\ee
In eqn. (\ref{poisson}) we have assumed for simplicity that the gauge 
kinetic function is a linear function of $\varphi$ with 
$B_F(\varphi)=1+\beta\varphi$.
Hence the variation of the fine structure constant away from the
string core is given by
\be
2 \pi r \frac{d \varphi}{dr}=  \beta I(r)  \mu (r_{\rm max})\,,
\ee
where 
\be
\mu(r)=2 \pi \int^r_0 \rho(r') r' dr'\,,
\ee
and 
\be
I(r)=\frac{\pi}{2\mu (r_{\rm max})} \int_0^r f^2(r') r' dr'\,,
\ee
is a function of $r$ smaller than unity. Here $r_{\rm max}$ represents a 
cut-off scale which is in a cosmological context of the order of the string 
correlation length. Far away from the string core $I(r)$ is a slowly 
varying function of $r$ which is always smaller than unity. An approximate 
solution for the behaviour of the field $\varphi$ may be obtained by taking 
$I(r) \sim {\rm const}$
\be
\varphi \sim  \frac{\beta I  \mu(r_{\rm max})}{2\pi} \ln \left(\frac{r}{r_0}\right)\,,
\ee
where $r_0$ is an integration constant. Since $\epsilon=B_F(\varphi)^{-1/2}$ 
we have 
\be
\epsilon \sim 1-\frac{\beta^2 I  \mu(r_{\rm max})}{4\pi} \ln \left(\frac{r}{r_0}\right)\,,
\label{epsilonr}
\ee
We see that the variation of the fine structure constant away from the
string core is proportional to the gravitational potential induced by
the strings. The value of $G \mu$ is constrained to be small
($\lapp 10^{-6}-10^{-7}$) in order to avoid conflict with CMB and LSS
results \cite{mu1,Wu1,Durrer,mu2,Wu2} (or even smaller depending on the 
decaying channels available to the cosmic string network \cite{mu3}). 
Note that these constraints are for standard local strings described by the 
Nambu-Goto action. Even though the cosmic strings studied in our paper 
are non-standard, having a local and a global component, we expect the 
limits on $G\mu(r_{\rm max})$ to be similar as those on $G \mu$ for 
standard local strings.
On the other hand the factor $\beta$ 
appearing in
eqn. (\ref{epsilonr}) is constrained by equivalence principle tests to be 
$|\beta| < 10^{-3} \, G^{1/2}$ \cite{OlivePos,Will,Damour}.  Hence, 
taking into 
account that we cannot observe scales larger than the horizon 
($\sim \, 10^4 \, {\rm Mpc}$) and are unlikely to 
probe variations of $\alpha$ at a distance much smaller than $1 \, {\rm pc}$ 
from a cosmic string, eqn. (\ref{epsilonr}) implies that a conservative 
overall limit on observable variations of $\alpha$ seeded by cosmic strings 
is $\Delta \alpha / \alpha \lapp 10^{-12}$ which is too small to have any 
significant cosmological impact. We thus see that, even 
allowing for a large contribution coming from the logarithmic factor in 
eqn. (\ref{epsilonr}), the spatial variations of $\alpha$ induced by such 
strings are too small to be detectable. In particular, this result means 
that despite the claim to the contrary in ref. \cite{Magueijo} cosmic 
strings will be unable to generate inhomogeneities in $\alpha$ which could 
be responsible for the generation of inhomogeneous reionization scenarios.

\section{Conclusion}

In this article we studied Nielsen-Olesen vortex solutions in 
Bekenstein-type models considering models with a generic gauge kinetic 
function. We showed that there is a class of models of this type for 
which the classical Nielsen-Olesen vortex is still a valid solution (with 
no $\alpha$ variation).
However, in general, spatial variations of $\alpha$ will be sourced by 
the electromagnetic energy concentrated along the string core. These  
are roughly proportional to the gravitational potential induced by the strings
which is constrained to be small. We have shown that
Equivalence Principle constraints impose tight limits on the allowed
variations of $\alpha$ on cosmological scales induced by
cosmic string networks of this type. As mentioned in ref. \cite{Magueijo} 
other defect
solutions may be considered such as monopoles or textures but in this case 
we expect similar conclusions to be drawn as far as the cosmological 
consequences of the theories are concerned. However, despite the claim in ref. 
\cite{Magueijo}
that domain walls can not be associated with changing-alpha theories
if the scalar field $\varphi$ is endowed with a potential
with a $Z_2$ symmetry then domain-walls may form and therefore
large scale inhomogeneities in the value of $\alpha$ associated with
different domains may be generated. We shall leave for future work a more
detailed study of other defect solutions and cosmological implications.

\ack
We thank Carlos Martins and Joana Oliveira for useful discussions.
J. Menezes was supported by a Brazilian Government grant - CAPES-BRAZIL 
(BEX-1970/02-0).  
Additional support came from Funda{\c c}\~ao para a Ci\^encia e a Tecnologia 
(Portugal) under contract POCTI/FNU/49507/2002-FEDER.

\end{document}